\documentclass[aip,pop,reprint,showpacs,superscriptaddress]{revtex4-1}
\usepackage{graphicx}
\usepackage{dcolumn}
\usepackage{bm}
\usepackage{color}
\usepackage{lipsum}
\usepackage{mathrsfs}
\usepackage{lineno}
\usepackage[colorlinks,urlcolor=blue,linkcolor=blue,citecolor=blue]{hyperref}
\usepackage{siunitx}
\begin{document}


\title{Head-on collision of large-scale high density plasmas jets: a first-principle kinetic simulation approach}
\author{D. Wu}
\email{dwu.phys@sjtu.edu.cn}
\affiliation{Key Laboratory for Laser Plasmas and School of Physics and Astronomy, and Collaborative Innovation Center of IFSA (CICIFSA), 
Shanghai Jiao Tong University, Shanghai 200240, People’s Republic of China}
\author{J. Zhang}
\email{jzhang@iphy.ac.cn}
\affiliation{Beijing National Laboratory for Condensed Matter Physics, Institute of Physics, Chinese Academy of Sciences, Beijing 100190, People’s Republic of China}
\affiliation{Key Laboratory for Laser Plasmas and School of Physics and Astronomy, and Collaborative Innovation Center of IFSA (CICIFSA), 
Shanghai Jiao Tong University, Shanghai 200240, People’s Republic of China}
\date{\today}

\begin{abstract}
In the double-cone ignition (DCI) inertial confinement fusion scheme, head-on collision of high density plasma jets is one of the most distinguished feature when compared with the traditional central ignition and fast ignition of inertial confinement fusion. However, the application of traditional hydrodynamic simulation methods become limited, due to serious plasma penetrations, mixing and kinetic physics that might occur in the collision process. To overcome such limitations, we propose a new simulation method for large-scale high density plasmas. This method takes advantages of modern particle-in-cell simulation techniques and binary Monte Carlo collisions, including both long-range collective electromagnetic fields and short-range particle-particle interactions. Especially, in this method, the restrictions of simulation grid size and time step, which usually appear in a fully kinetic description, are eliminated. In addition, collisional coupling and state-dependent coefficients, that are usually approximately used with different forms in fluid descriptions, are also removed in this method. Energy and momentum exchanges among particles and species, such as thermal conductions and frictions, are modelled by ``first principle'' kinetic approaches. The correctness and robustness of the new simulation method are verified, by comparing with fully kinetic simulations at small scales and purely hydrodynamic simulations at large scale. Following the conceptual design of the DCI scheme, the colliding process of two plasma jets with initial density of 100 g/cc, initial thermal temperature of 70 eV, and counter-propagating velocity at 300 km/s is investigated using this new simulation method. Quantitative values, including density increment, increased plasma temperature, confinement time at stagnation and conversion efficiency from the colliding kinetic energy to thermal energy, are obtained with a density increment of about three times, plasma temperature of 400 eV, confinement time at stagnation of $50$ ps and conversion efficiency of $85\%$. These values agree with the recent experimental measurements at a reasonable range.

\end{abstract}

\pacs{52.38.Kd, 41.75.Jv, 52.35.Mw, 52.59.-f}
\maketitle

\section{Introduction}
As a promising candidate for inertial confinement fusion (ICF) ignition \cite{icf1, icf2,icf3}, the double-cone ignition
(DCI) scheme \cite{icf4} was proposed by J. Zhang recently. The scheme is
composed of four progressive controllable processes: quasi-isentropic compression, acceleration, head-on collision and rapid heating of the compressed DT fuel.
The DT fuel shells in gold-cones are firstly compressed and accelerated by driving laser pulses along the axis of the cones, forming supersonic plasma jets. The plasma jets then collide at the center of the open space between the two cones, forming a preheated plasma with an increased density. The preheated plasma is further rapidly heated to the conditions required for thermonuclear ignition by strong beams of MeV electrons generated by picosecond, petawatt heating laser pulses. As one of the main processes in the DCI scheme, head-on collision and subsequent evolution of counter-streaming plasmas also occurs in many other laser plasma researches  \cite{hol1, hol2}, such as x-ray lasers employing laser-irradiated targets and plasma blow-off inside hohlraums in indirect-drive ICF researches.

In reality, the head-on collision of plasma jets spans a wide range of Coulomb collisions. However, a simple fluid description is not always valid in collisions of  high-density plasmas. As a single fluid model allows for only one value of the flow velocity at a given spatial location, interpenetration of the plasmas is prohibited. Such a limitation of single fluid model results in immediate stagnation of plasma jets with complete conversion of kinetic energies to thermal ones. This limitation can be eliminated by employing a multiple-ion fluid description, where each separate plasma jet is described as a collisional fluid, with collisional coupling with each other through velocity and state-dependent coeﬀicients. However as the colliding plasma jets meet at high relative velocity with low density fronts, the collisionality therein can be very low with ion-ion mean-free paths comparable to the system size. The validity of multiple ion fluid de scription becomes suspect as the plasmas heat and overlap in velocity space. A kinetic-ion model is therefore required. In addition, a kinetic-ion model with ion distributions to be non-Maxwellian allows to intrinsically capture many effects that are difficult or impossible to capture in a fluid-ion description. Beside the interpenetration of plasma jets in head-on collisions, phenomena like ion-viscosity, ion-acceleration, finite ion gyroradii effects, nonisotropic pressure, diffusion and kinetic mixing at shocked interfaces, and nonthermal fusion production may also require a kinetic-ion treatment.

Nowadays, fully kinetic model treating both electrons and ions kinetically is widely used and available in plasma physics communities. Although it is the most accurate model, electron spatial scales, such as the electron Debye length or electron gyroradius, and their associated time scales must be resolved \cite{pic}. As for the head-on collision in the DCI scheme, the typical values of density, size and duration of the plasma jet are  $\sim100$ g/cc, $\sim100$ $\mu$m and $\sim100$ ps, respectively. The required simulation grids and time steps are enormous.
Thus, it is by no means possible to model such a large spatial/temporal scale high density plasma by using a fully kinetic method.

In this paper, we propose a new simulation method for large-scale and high density plasmas, with an ingenious kinetic-ion and kinetic/hydrodynamic-electron treatment. 
This method takes advantages of modern particle-in-cell (PIC) simulation techniques and binary Monte Carlo (MC) collisions, which include both long-range collective electromagnetic fields and short-range particle-particle interactions. The correctness and robustness of the new simulation method are verified, by comparing with fully kinetic simulations at small scales and purely hydrodynamic
simulations at large scale.Following the conceptual design of the DCI scheme, the colliding process of two plasma jets with initial density of 100 g/cc, initial thermal temperature of 70 eV, and counter-propagating velocity at 300 km/s is investigated using this new simulation method. Quantitative values, including density increment, increased plasma temperature, confinement time at stagnation and conversion efficiency from the colliding kinetic energy to thermal energy, are obtained with a density increment of about three times, plasma temperature of 400 eV, confinement time at stagnation of $50$ ps and conversion efficiency of $85\%$. These values agree with the recent experimental measurements at a reasonable range.
 
The paper is organized as follows. Details of the simulation model and algorithm are described in section-II. Benchmark of the simulation code is displayed in section-III, where we have compared with both fully kinetic method at small scales and hydrodynamic method at large scales. A simulation run for a head-on collision of high density plasma jets using practical parameters of the DCI is presented and analysed in Section-IV. Finialy, discussions and summaries are made in Section-V.

\begin{figure*}
	\includegraphics[width=15.0cm]{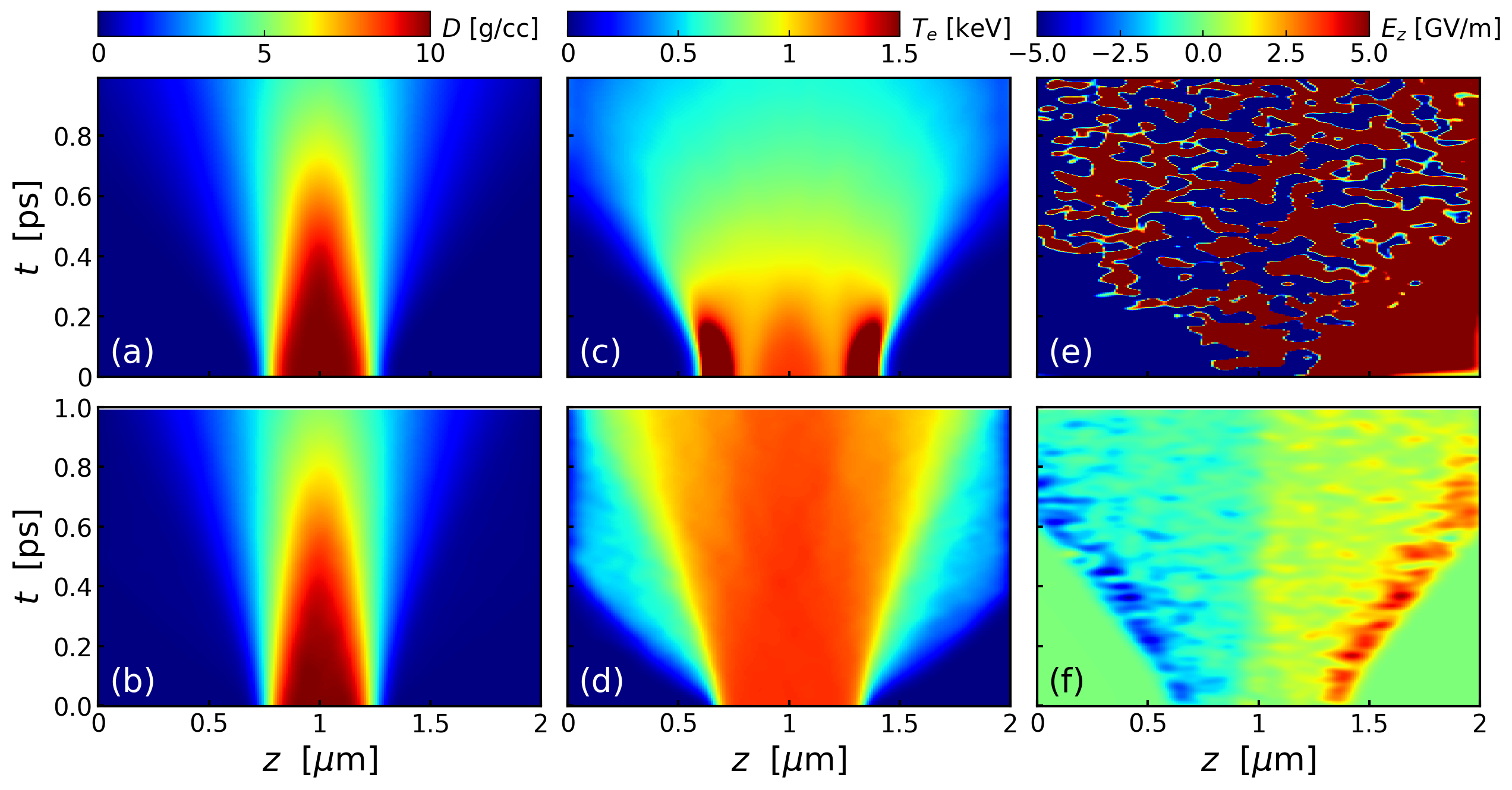}
	\caption{\label{fig1}(color online)
		The spatial-temporal evolutions of plasma density, electron-temperature and electric fields: (a), (c) and (e) are obtained by fully kinetic simulation method, 
		and (b), (d) and (f) are obtained by the KIKFE method.}
\end{figure*}

\begin{figure*}
	\includegraphics[width=15.0cm]{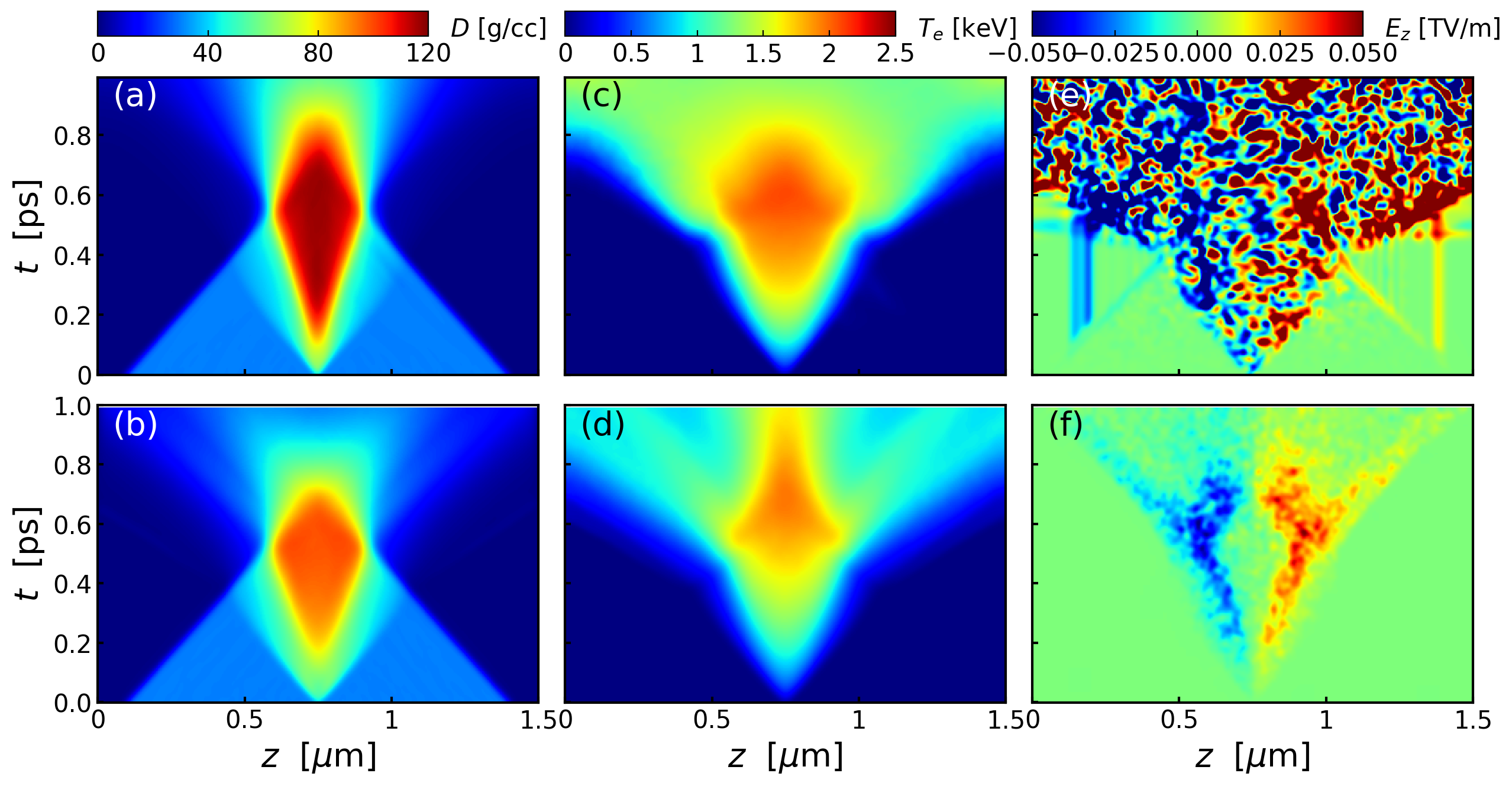}
	\caption{\label{fig2}(color online)
		The spatial-temporal evolutions of plasma density, electron-temperature and electric fields: (a), (c) and (e) are obtained by fully kinetic simulation method, 
		and (b), (d) and (f) are obtained by the KIKFE method.}
\end{figure*}

\begin{figure}
	\includegraphics[width=8.5cm]{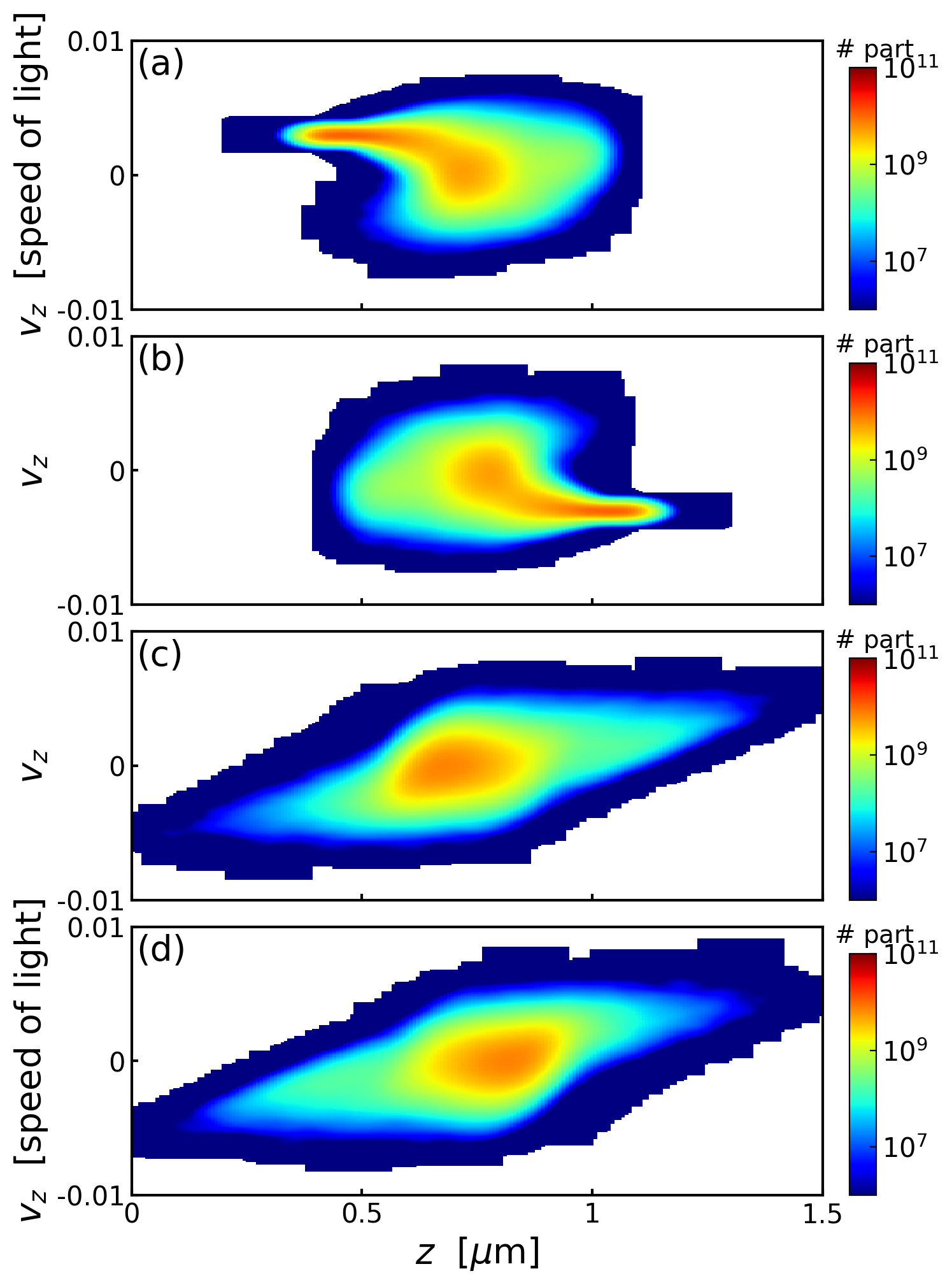}
	\caption{\label{fig3}(color online) (a)-(b) show the $v_z$-$z$ phase space distributions at $0.33$ ps, for jet-1 and jet-2 respectively. (c)-(d) show the $v_z$-$z$ phase space distributions at $0.66$ ps, for jet-1 and jet-2 respectively. The simulation data are obtained by using the KIKFE method, and results obtained by the fully kinetic method are similar.}
\end{figure}

\begin{figure}
	\includegraphics[width=8.5cm]{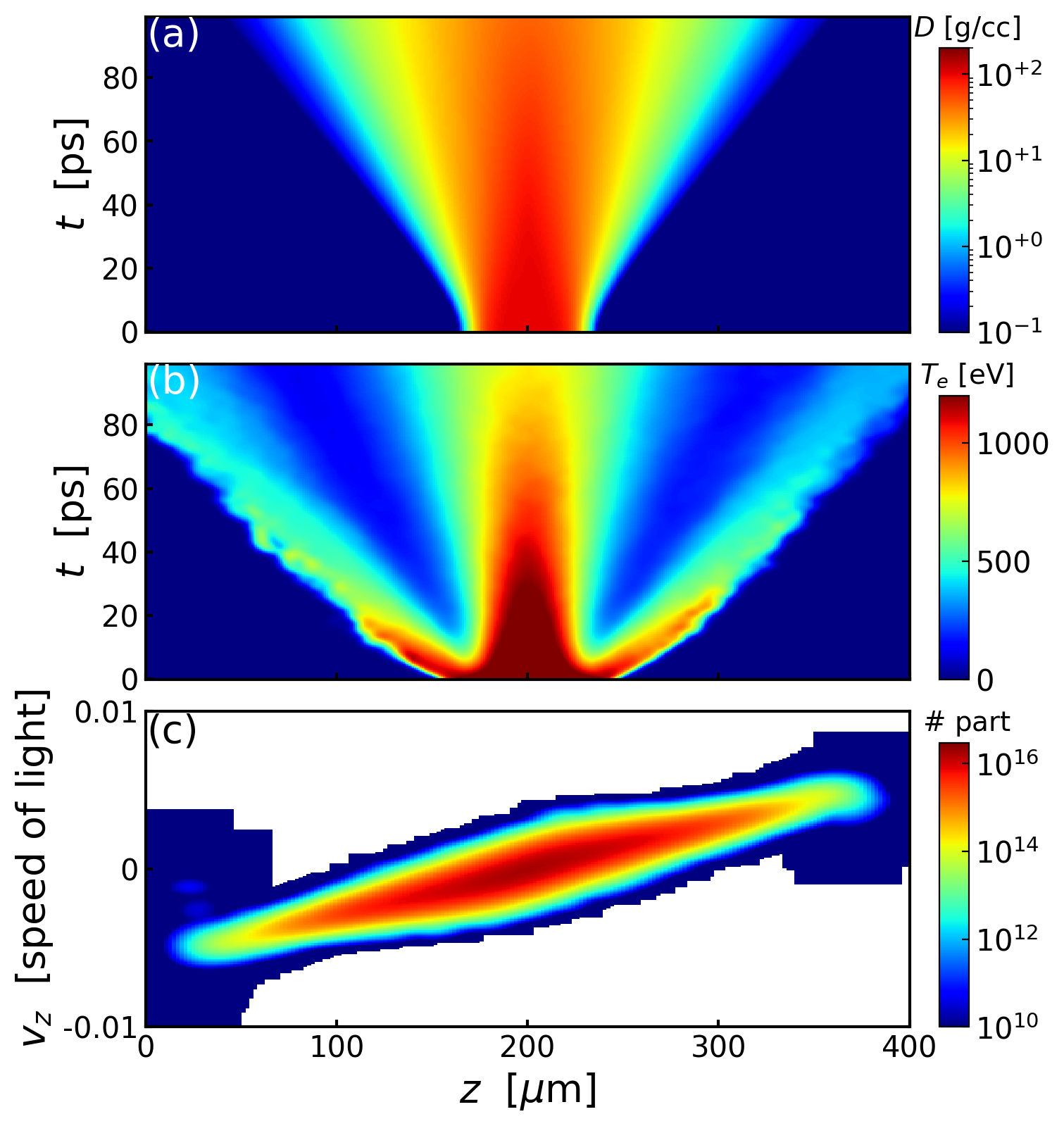}
	\caption{\label{fig4}(color online) (a)-(b) show the spatial-temporal evolutions of plasma density and electron-temperature, (c) the phase-space plot of Langrangian-ion particles at $100$ ps. The simulation data are obtained by using the KIKFE method.}
\end{figure}

\section{Algorithm of the model}

We start from the rigorous fully kinetic Fokker-Planck model with $f_{i}(\textbf{r},\textbf{v},t)$ and $f_{e}(\textbf{r},\textbf{v},t)$ describing the phase space distributions of ions and electrons,

\begin{equation}
	\label{eq1}
	\frac{\partial f_i}{\partial t}+\textbf{v}\cdot\frac{\partial f_i}{\partial \textbf{r}}-\frac{q_i}{m_i}(\textbf{E}+{\textbf{v}\times\textbf{B}})\cdot\frac{\partial f_i}{\partial \textbf{v}}=\frac{\partial f_{i}}{\partial t}|_{\text{i,i; i,e}}, \nonumber
\end{equation}

\begin{equation}
	\label{eq2}
	\frac{\partial f_e}{\partial t}+\textbf{v}\cdot\frac{\partial f_e}{\partial \textbf{r}}+\frac{q_e}{m_e}(\textbf{E}+{\textbf{v}\times\textbf{B}})\cdot\frac{\partial f_e}{\partial \textbf{v}}=\frac{\partial f_{e}}{\partial t}|_{\text{e,e; e,i}}. \nonumber
\end{equation}
On the right hand sides of Eqs. (\ref{eq1}), ${\partial f}/{\partial t}|_{\text{i,i; i,e}}$ represents the close collisions with ions and electrons separately. The collision operator is of the form,  
\begin{eqnarray}
	\frac{\partial f}{\partial t}|_{\text{c}}=&&
	\int d^3p_2 \int d^3p_3 \int d^3p_4 W(\textbf{p}_1, \textbf{p}_2; \textbf{p}_3, \textbf{p}_4) \nonumber \\
	&&\times\{f(\textbf{r}, \textbf{p}_3)f(\textbf{r}, \textbf{p}_4)-
	f(\textbf{r}, \textbf{p}_1)f(\textbf{r}, \textbf{p}_2)\},
\end{eqnarray}
where $W(\textbf{p}_1, \textbf{p}_2; \textbf{p}_3, \textbf{p}_4)$ is the collision rate, which is a function of plasma temperature and density.
The integral takes into account all possible scatterings: the two particles with initial momenta ($\textbf{p}_1^{\text{in}}$, $\textbf{p}_2^{\text{in}}$) scatter into the finial momenta ($\textbf{p}_3^{\text{out}}$, $\textbf{p}_4^{\text{out}}$); and by symmetry processes starting with ($\textbf{p}_3^{\text{in}}$, $\textbf{p}_4^{\text{in}}$) and ending in ($\textbf{p}_1^{\text{out}}$, $\textbf{p}_2^{\text{out}}$). For high density plasmas, when the quantum degenerate effects are not ignorable, Pauli exclusion principle must be taken into account. 
For those collisions, a more general collision operator reads,   
\begin{eqnarray}
	\frac{\partial f}{\partial t}|_{\text{c}}=&&\int d^3p_2 \int d^3p_3 \int d^3p_4 W(\textbf{p}_1, \textbf{p}_2; \textbf{p}_3, \textbf{p}_4)\nonumber \\
	&&\times \{ f(\textbf{r}, \textbf{p}_3)f(\textbf{r}, \textbf{p}_4)[1-f(\textbf{r}, \textbf{p}_1)][1-f(\textbf{r}, \textbf{p}_2)]-\nonumber \\
	&& f(\textbf{r}, \textbf{p}_1)f(\textbf{r}, \textbf{p}_2)[1-f(\textbf{r}, \textbf{p}_3)][1-f(\textbf{r}, \textbf{p}_4)] \},
\end{eqnarray}
where the Pauli exclusion principle shows up explicitly here \cite{pauli1,pauli2}, imposing that $f$ should be less than $1$.
Namely, no more than one fermion can occupy a phase space cell of volume $(2\pi\hbar)^3$.

The evolution of phase space distributions are governed by both collective electromagnetic fields and close collision interactions. 
Here, the electromagnetic fields are governed by Maxwell Equations, which read,
\begin{equation}
	\partial \textbf{E}/\partial t=-2\pi(\textbf{J}_i+\textbf{J}_e)+\nabla\times\textbf{B}, \nonumber
\end{equation}

\begin{equation}
	\label{eq6}
	\partial \textbf{B}/\partial t=-\nabla\times\textbf{E}, \nonumber
\end{equation}
here $\textbf{J}_{i,e}$ are current densities of ions or electrons, which are defined by 

\begin{equation}
	\label{eq7}
	\textbf{J}_{i,e}=\int \textbf{v}_{i,e} f_{i,e}d\textbf{v}_{i,e}. \nonumber
\end{equation}

The above coupled Eqs.\ (\ref{eq1})-(\ref{eq7}) make a complete description of plasmas. One of the efficient ways for solving the above coupled equations is the PIC method, which is an Euler-Lagrange approach. When compared with the finite difference method used in a Fokker-Planck code, the use of Langrangian particles in phase space has the advantage of naturally resolving the large colliding velocities, without the difficulty of meshing a large region of velocity space, which is necessary in a finite difference Fokker-Planck code. 
Although rigorous, complete and can achieved via a PIC method, the above model is only applicable for small scale plasmas, as the electron spatial and temporal scales must be resolved. 

In the quasi-neutral and current-free limit, restrictions on electron spatial and temporal resolutions in the fully kinetic model can be eliminated. This is indeed the case for the head-on collision of high density plamas, as the electron spatial and temporal scales are much smaller than the system spatial size and temporal duration. Electron dynamics can now be described only by macroscopic quantities,
\begin{equation}
\label{eq8}
\frac{d \textbf{u}_e}{dt}= \frac{q_e}{m_e}(\textbf{E}+\textbf{u}_e\times\textbf{B})-\frac{\nabla p_{e}}{m_en_e}+\sum_{i}\mu_{e,i}(\textbf{u}_e-\textbf{v}_{i}). \nonumber
\end{equation}
In Eq.\ (\ref{eq8}), $n_e$ and $\textbf{u}_e$ are the density and velocity of electrons, $p_e$ is the electron pressure, and $\mu_{e,i}$ is the collision frequency of electrons with ions. The quasi-neutral and current-free conditions also imply that the inertia of electrons can be ignored, which indicate that the electric field can be solved by 

\begin{equation}
	\label{eq9}
	\textbf{E}=-\textbf{u}_e\times\textbf{B}+\frac{\nabla p_{e}}{q_en_e}-\frac{m_e}{q_e}\sum_{i}\mu_{e,i}(\textbf{u}_e-\textbf{v}_{i}).
\end{equation}
In the quasi-neutral and current-free limits, $n_e$ and $\textbf{u}_e$ are therefore defined as 
\begin{eqnarray}
	n_e&\equiv&\sum_i Z_i n_i \nonumber \\
	n_e\textbf{u}_e&\equiv&\sum_i Z_in_i \textbf{v}_i,
\end{eqnarray}
with $n_i$, $\textbf{v}_i$ and $Z_i$ the density, velocity and charge state of ions, respectively. In the code, $n_e$ and $\textbf{u}_e$ are obtained on the spatial grids via appropriate interpretations over the surrounding Langrangian particles. 

In Eq.\ (\ref{eq9}), the electric field is determined by three contributions, from Hall effect, electron pressure gradient, and electron-ion collisions respectively. 
The contribution to the electric field due to Hall effect, $\textbf{u}_e\times\textbf{B}$, is easy to handle, as $\textbf{B}$ can be solved explicitly via Eq.\ (\ref{eq6}). 

The contribution due to electron-ion collisions can be rewritten as 
\begin{eqnarray}
	\frac{m_e}{q_e}\sum_{i}\mu_{e,i}(\textbf{u}_e-\textbf{v}_{i})&=&\frac{1}{q_en_e}\sum_{i}m_{i}n_{i}\mu_{i,e}(\textbf{u}_e-\textbf{v}_{i}) \nonumber \\
	&\equiv&\frac{1}{q_en_e}\sum_{i}m_{i}n_{i}\frac{\partial \textbf{v}_{i}}{\partial t}|_{i, e},
\end{eqnarray}
where the relationship $m_en_e\mu_{e,i}=m_in_i\mu_{i,e}$ is considered. 
Furthermore, by replacing the frictional rate of change due to electron collisions with the actual change in a time step, this part of electric field becomes,
\begin{equation}
	\label{eq12}
	\frac{1}{q_en_e}\sum_{i}m_{i}n_{i}\frac{\partial \textbf{v}_{i}}{\partial t}|_{i, e}\equiv\frac{1}{q_en_e}\sum_{i}m_in_i(\frac{\Delta \textbf{v}_i}{\Delta t})|_{i,e}.
\end{equation}
Eq.\ (\ref{eq12}) can be solved via appropriate interpretations over the Langrangian ions before and after collisions with electrons in one time step. 

As for the contribution due to electron pressure gradient, one needs to first define the electron pressure and then obtain the evolution of which. 
Generally, the drift kinetic energy of electrons is much smaller than their thermal kinetic energy. For example, with drift velocity of $300$ km/s for head-on collisions of plasma jets, the drift kinetic energy of electrons is only $0.3$ eV.
Thus, the electron pressure is equivalent to the electron kinetic energy density with good approximations.
We therefore have 
\begin{equation}
p_e(\textbf{r}, t)=\int m_ev^2 f_e(\textbf{r}, \textbf{v}, t) d\textbf{v}.
\end{equation}
When ignoring radiation processes, the evolution of electron kinetic energy density is mainly determined by the convections, Ohm's heating, electron-ion energy exchange and thermal conductions, respectively,
\begin{eqnarray}
	\label{eq14}
	\frac{\partial p_e(\textbf{r}, t)}{\partial t} =&-&\nabla\cdot(p_e\textbf{u}_e)+q_en_e\textbf{u}_e\textbf{E} \nonumber \\
	&+&n_e\mu_{ei}(T_i-T_e)+\nabla\cdot(\kappa\nabla T_e).
\end{eqnarray}
We do not solve the above equation of pressure evolution via finite different method. Instead, this equation is solved via some new concepts. 
As for the convection term, $\nabla\cdot(p_e\textbf{u}_e)$, it is simulated by the movement of each Langrangian electron-particle, with $\textbf{r}^{\text{new}}_{\alpha}=\textbf{r}^{\text{old}}_{\alpha}+\textbf{v}_e dt$, where $\alpha$ is the index for Langrangian electron particles, and $\textbf{v}_e=\textbf{u}_e+\tilde{\textbf{v}}_e$, including a comment moving velocity defined as $\textbf{u}_e\equiv\sum_i Z_in_i \textbf{v}_i/n_e$ and a fluctuation velocity defined by local temperatures.
Please note the definition of moving velocity $\textbf{u}_e$ is also the requirement of quasi-neutral and current-free conditions.
As for the Ohm's heating term, $q_en_e\textbf{u}_e\textbf{E}$, it is achieved by numerical heating or cooling each Langrangian electron-particle within a same computational cell if $q_en_e\textbf{u}_e\textbf{E}$ is positive or negative. 
Finialy for terms of electron-ion energy exchange and thermal conductions, $n_e\mu_{ei}(T_i-T_e)+\nabla\cdot(\kappa\nabla T_e)$, they can be well handled by binary MC collisions. The binary MC particle-particle collision model is essentially that of Takizuka \cite{Taki}, Nanbu \cite{Nanbu}, and Sentoku \cite{Sentoku} but with the inclusion of Pauli-exclusion principles \cite{pauli1, pauli2}.  
At each time step, Langrangian particles within a same computational cell are randomly paired up and do scattering operations.
Since the scatter of each pair is kinematically correct, the method rigorously conserves momentum and energy for equally weighted particles. Extensive benchmarks have already verified that this method is equivalent to a Fokker-Planck description of Coulomb scattering in the limit of small angular collisions.
Therefore, the collisional coupling and state-dependent coefficients are no longer needed as used in fluid descriptions, as such coupling coefficients are usually obtained with approximations of the Fokker-Planck collision operator. Note however in a fluid model, 
such coupling coefficients are not even uniquely defined \cite{fl1, fl2, fl3,fl4}, with various researchers employing different approximations. 

With the new electric and magnetic fields obtained via Eq.\ (\ref{eq9}) and Eq.\ (\ref{eq6}), 
the ions particle trajectories are advanced in phase space, according to
\begin{eqnarray}
	\frac{d \textbf{r}_i}{dt}&=&\textbf{v}_i \nonumber \\
	\frac{d \textbf{v}_i}{dt}&=&\frac{q_i }{m_i}\textbf{E}+\frac{q_i }{m_i}\textbf{v}_i\times\textbf{B}+\frac{\partial \textbf{v}_i}{\partial t}|_{i,e}+\sum_{j}\frac{\partial \textbf{v}_i}{\partial t}|_{i,j}.
\end{eqnarray}
The Coulomb collisions include interactions with both electrons and all ions as denoted by the sum over, which can be also calculated by binary MC collisions. 

As a brief summary, in this new method, the ion kinetic effects are fully retained. 
For electrons, although they are still discretized with Lagrangian particles, their macroscopic dynamics are modelled by using a fluid approach. 
Therefore,
the resolutions of electron spatial scales and their associated time scales are no longer needed, which is vital for the simulations of large scale high density plasmas.
As both electrons and ions are discretized with Lagrangian particles, 
and the MC collision methods are used to model close particle-particle interactions, state-dependent coupling coefficients as used in hydrodynamic models are therefore no longer required.

\section{Benchmark of the model}
This new simulation method, i.e., the kinetic-ion kinetic/hydrodynamic electron (KIKFE) method, has already been achieved and implemented into the LAPINS \cite{lapins1,lapins2} code. 
In order to verify the robustness and correctness of this method, comparisons with both fully kinetic method at small scales and hydrodynamic method at large scales are taken in this section.

\subsection{Comparison with fully kinetic simulations}
The free expansion of a deuterium plasma slab with density $10$ g/cc, size $0.4$ $\mu$m and temperature $1300$ eV is simulated by using both a fully kinetic simulation method and the proposed KIKFE method. In both methods, the simulation grid size, time step and particles in a computational cell are kept the same, with $dz=0.01$ $\mu$m, $dt=0.166$ fs and each computational cell is filled with $2000$ Langrangian particles of electrons and ions. In Fig.\ \ref{fig1}, the spatial-temporal evolutions of plasma density, ion-temperature and electric fields are displayed. 
Fig.\ \ref{fig1}(a) and (b) display the evolution of plasma density calculated by using the fully kinetic and the KIKFE methods, respectively.
As one can see, the evolution of plasma density coincides with each other quite well. 
Fig.\ \ref{fig2}(c) and (d) display the evolution of electron-temperature. 
In the simulations, the electron-temperature is calculated in each computational cell by counting all Lagrangian electrons.
As one can see, the evolution of plasma electron-temperature also coincides with each other quite well. However we noticed that KIKFE method can not reproduce the highly localized temperature distributions on the plasma-vacuum boundary at the initial expansion stage. This might due to the fact that charge-separation electric field is not included in our KIKFE method. On small scale simulations, charge-separation electric field usually plays an important role in accelerating and heating charged particles.        
In Fig.\ \ref{fig1}(e) and (f), the evolution of electric fields calculated by using the fully kinetic and the KIKFE methods are also compared with each other. 
As one can see,  in Fig.\ \ref{fig1}(e), there exists significant electric perturbations, which often appear in fully kinetic simulations. 
However, as for the electric fields in Fig.\ \ref{fig1}(f), which are calculated by the KIKFE method, such perturbations disappear. 
The significantly clearer electric field distributions demonstrate the advantages of KIKFE method in simulating high density plasmas.

\begin{figure}
	\includegraphics[width=8.5cm]{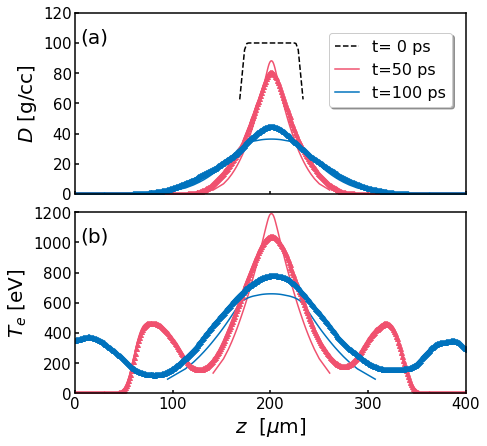}
	\caption{\label{fig5}(color online) The plots of plasma density and electron temperature as a function of time: dashed lines are the initial density, 
		solid and triangles lines are obtained by the hydrodynamic and the proposed KIKFE simulation methods, separately; 
		red and blue colors represent different data values at $t=50$ and $t=100$ ps, respectively.}
\end{figure}

\begin{figure*}
	\includegraphics[width=15.0cm]{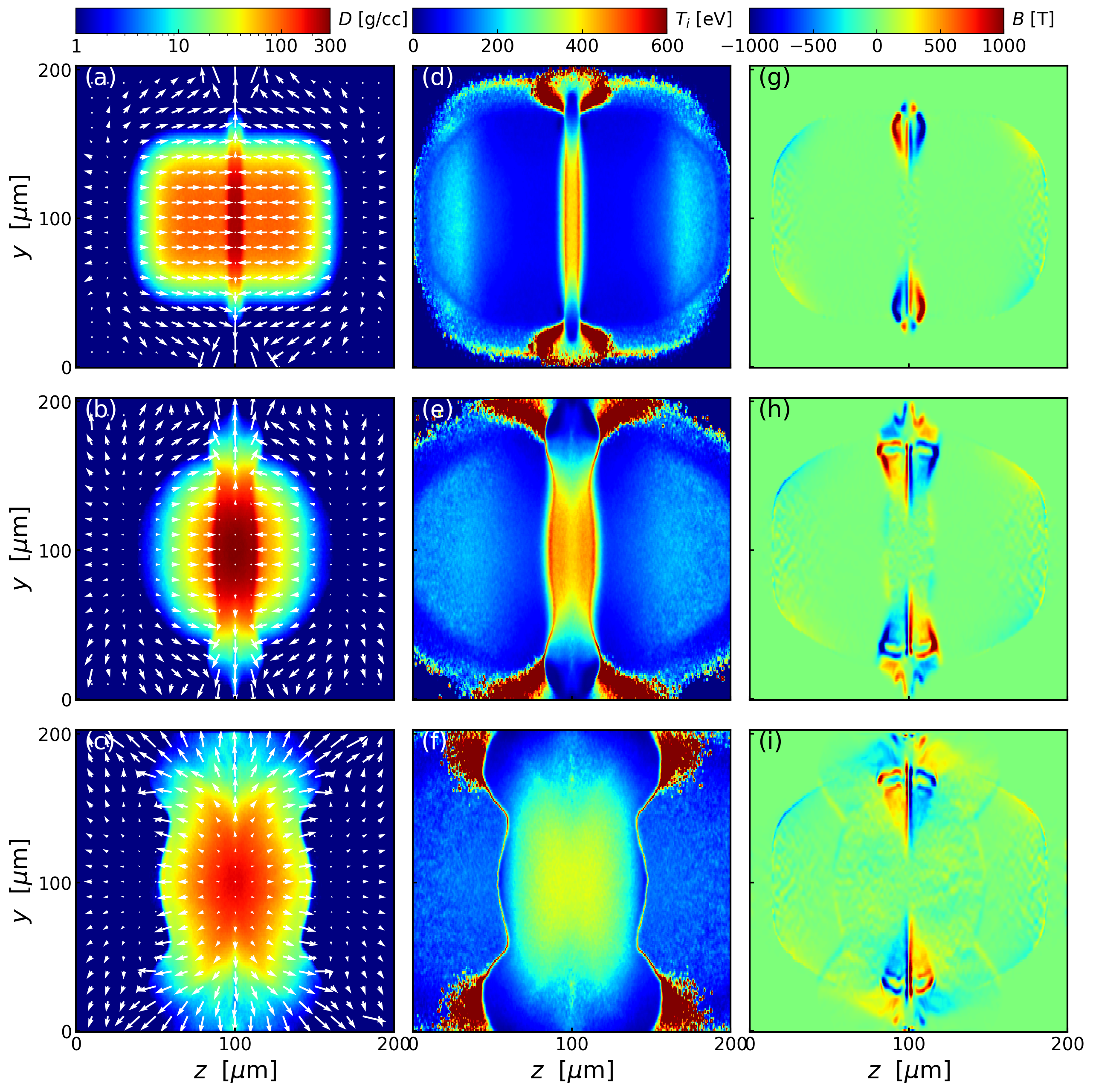}
	\caption{\label{fig6}(color online)
		The evolutions of plasma density (1-st row), ion temperature (2-nd row) and magnetic fields (3-rd row) at $t=66$ (1-st line), $t=231$ ps (2-nd line) and $333$ ps (3-rd line), respectively.
		The white arrows indicate the magnitude and direction of the averaged ion-velocity.}
\end{figure*}

\begin{figure*}
	\includegraphics[width=15.5cm]{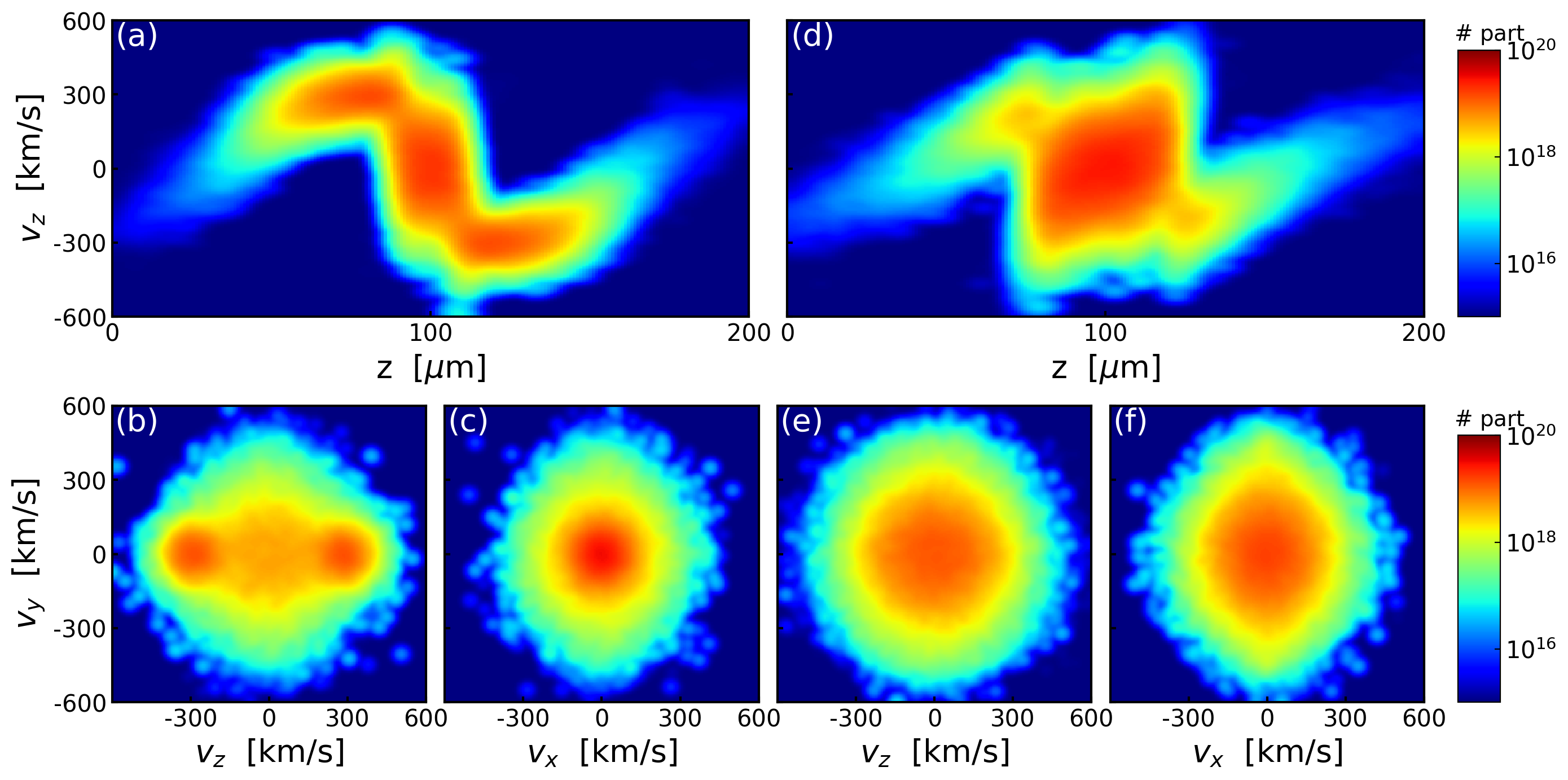}
	\caption{\label{fig7}(color online)
		The phase-space plots of ions at $t=100$ ps [(a)-(c)] and $t=166$ ps [(d)-(f)], respectively.}
\end{figure*}

\begin{figure}
	\includegraphics[width=8.5cm]{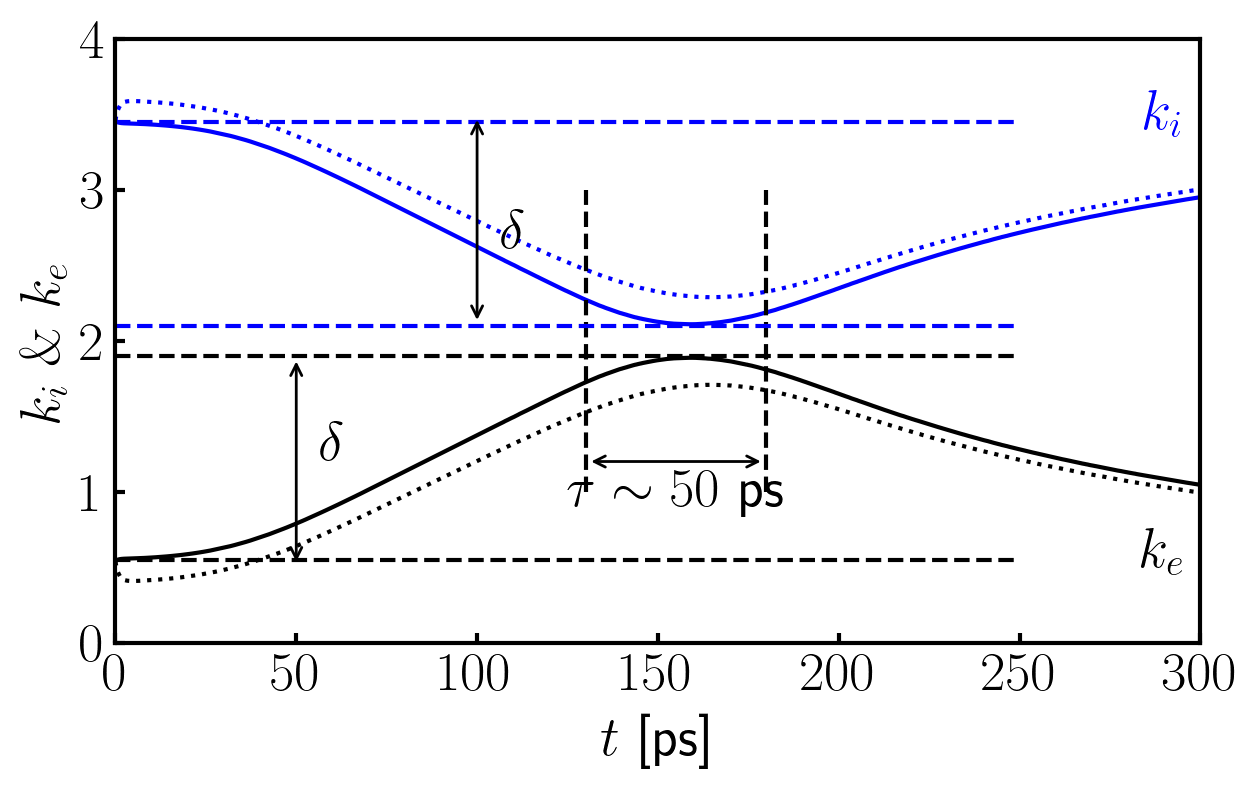}
	\caption{\label{fig8}(color online)
		The temporal evolution of global kinetic energies for electrons and ions, with arbitrary unit used in the plot.
		Solid lines display temporal evolution of colliding process for two degenerate plasma jets with initial density of 100 g/cc, temperature of 65 eV and drifting velocity of 300 km/s, and dashed lines display the results of a toy model in which electrons are treated as classical particles. 
		The window marked between two black dashed lines indicate the confining window at stagnation for the colliding plasmas, which is about $50$ ps for the considered cases.
		See text for explanations of $\delta$.}
\end{figure}

The colliding of two plasma jets at small scales are also benchmarked in this section. Two head-on colliding plasmas jets with density of $30$ g/cc, spatial size of $0.5$ $\mu$m and colliding velocity of $900$ km/s  are simulated by using both fully kinetic and the proposed KIKFE methods. In both methods, the simulation parameters, such as grid size, time step and particles in a computational cell, are also kept the same, with $dz=0.01$ $\mu$m, $dt=0.166$ fs and each computational cell is filled with $2000$ Langrangian particles of electrons and ions.
In Fig.\ \ref{fig2}, the spatial-temporal evolution of plasma density, electron temperature and electric fields are displayed. 
For colliding plasmas, the density compression ratio is one of the most concerned in the colliding regions. In Fig.\ \ref{fig2}(a) and Fig.\ \ref{fig2}(b), the density increment is quite similar with both simulation methods, which is about four times the initial colliding density. In Fig.\ \ref{fig2}(c) and (d), evolutions of electron temperatures are displayed. High temperature distributions appear at the colliding regions, which means the colliding kinetic energies are significantly converted into thermal ones therein. In Fig.\ \ref{fig2}(e) and (f), evolution of electric fields are displayed. Once again, significant perturbations appear when a fully kinetic simulation method is used, and for the KIKFE method, such perturbations also disappear. 

For colliding plasmas, plasma penetration and mixing are expected to occur, which is usually beyond the capability of traditional hydrodynamic simulations. In order to clearly figure out such effects, in our simulation, jet-1 (with colliding velocity along z direction) and jet-2 (with colliding velocity along -z direction) are marked with different tracers. In Fig.\ \ref{fig3}, phase space plots of plasma ions are displayed. In Fig.\ \ref{fig3}(a) and (b), ions of jet-1 and ions of jet-2 at $t=0.33$ ps are displayed respectively. Mutual penetration and mixing start to take place at the colliding regions. In Fig.\ \ref{fig3}(c) and (d), ions of jet-1 and ions of jet-2 at $t=0.66$ ps are displayed respectively, and such penetrations and mixing become more significant. 

\subsection{Comparison with hydrodynamic simulations}
In this benchmark, we increase the density and size of the deuterium plasma to $100$ g/cc and $40$ $\mu$m, and keep the initial temperature still at $1300$ eV.
Modelling the free expansion of such a large-scale high density plasma is almost impossible by using a fully kinetic method. 
However by using the proposed KIKFE method, the restrictions of simulation grid size and time step can be eliminated. 
In our simulation, grid size and time step are set to be $dz=1$ $\mu$m, $dt=16.6$ fs, and each computational cell is also filled with $2000$ Langrangian particles of electrons and ions. 
Fig.\ \ref{fig4}(a)-(b) show the evolution of plasma density and electron-temperature. 
Although the plasma density and electron-temperature are decreasing along the expansion fronts, we do have noticed peak values for electron-temperatures at the expanding low density fronts. 
At the expanding low density fronts, there exists peak values for electron temperature distributions. The existence of these peaks might due to the following two reasons.
Firstly, for the expanding fronts, the plasma density is very low and the distribution of ions significantly departs from Maxwell distributions. 
Secondly, as shown in Fig.\ \ref{fig1}(e) and (f), we do have found net accelerating electric fields located at the expanding fronts. 
This electric field is able to accelerate those ions and leads to a high electron temperature distribution therein.

In order to verify the correctness of KIKFE method at large scales, the hydrodynamic code MULTI-1D \cite{multi} is taken for the large scale benchmark of free expansion simulations.
The comparison between the LAPINS code with KIFKE method and the MULTI-1D code is displayed in Fig.\ \ref{fig5}.  
Here, Fig.\ \ref{fig5}(a) presents the changing of plasma density at $t=50$ and $t=100$ ps respectively. The black dashed line is the initial density profile. 
Solid lines display data calculated by the MULTI-1D code, and triangles display data by the LAPINS code. As one can see, at $t=50$ and $t=100$ ps, 
the calculated density profiles with different codes coincide with each other quite well. 
At $t=100$ ps, departures at the central regions appear, but still with a good agreement at low density fronts.
Fig.\ \ref{fig5}(b) displays the changing of electron-temperature at $t=50$ and $t=100$ ps, respectively. 
As one can see, electron-temperatures, especially the changing profiles, have a considerable agreement with each other.
A significant difference between LAPINS and MULTI-1D simulations lies at the low density fronts, 
with the former existing peak values and the latter decreasing smoothly along the expansion directions.
It is reasonable to think such departures actually reflect the advantages of the LAPINS code. As this can be well understood by recalling the net accelerating electric fields at the low density fronts.
The acceleration of ions located at the low density fronts can be clearly depicted at phase plots.
Fig.\ \ref{fig4}(c) shows the phase space plots of ion particles at $t=100$ ps calculated by the LAPINS code, which contains the entire ion-kinetic effects. 
It clearly shows that the departure of maxwell distribution at low density fronts is significant. At such regions, the fluid description certainly breaks down.

\section{The colliding of large scale and high energy density plasmas jets}

In the DCI scheme, the head-on colliding of high density plasma jets is of vital importance, 
as the aim of this colliding is to form a preheated plasma with an increased density. 
Following the DCI conceptual design, the density and velocity of the plasma jet is chosen as $100$ g/cc and $300$ km/s.
In this section, by using the newly developed LAPINS code, the colliding of such high density plasma jets is investigated. 
Quantitative values, like density increment, pre-heated plasma temperature, confinement time at stagnation and conversion efficiency from the colliding kinetic energy to thermal energy, are obtained in this investigation, which might serve as a reference for the future detailed studies. 

Here, the colliding of two plasma jets with an initial density $100$ g/cc, temperature $65$ eV and colliding velocity $300$ km/s are modelled 
within a two-dimensional in space and three-dimensional in velocity (2D-3V) framework. 
In the simulation, grid size and time step are set to be $dz=dy=1$ $\mu$m, $dt=16.6$ fs, and each computational cell is filled with $400$ Langrangian particles of electrons and ions.
As the thermal temperature is much smaller than the Fermi energy of plasma at density of 100 g/cc, the role of quantum degeneracy is not ignorable.   
Therefore, electrons are initialled with Fermi-Dirac distributions, and Pauli exclusion principle is taken into account for the evolution of electrons.

Figure\ \ref{fig6} display the evolution of plasma density, ion temperature and magnetic field, with white arrows indicating the magnitude and direction of averaged ion velocity, at $t=66$ ps, $t=231$ ps and $t=333$ ps, respectively.
The density evolution indicates that, at $t=66$ ps as shown in Fig.\ \ref{fig6}(a) when the low density fronts meet, penetration with each other occurs. 
At $t=231$ ps as shown in Fig.\ \ref{fig6}(b), the density increment occurs. 
Simulation shows that the density increment can be as high as three times the initial density. 
The evolution of ion temperature is displayed in Fig.\ \ref{fig6}(d)-(f).
As one can see at the initial stage $t=66$ ps, the majority of ions move at constant velocities of $300$ km/s along z directions. 
When the high density parts of plasmas meet, at $t=231$ ps, the deceleration and deflection of ion particles take places.
Such deceleration and deflection mean that some of ``well-ordered'' ion kinetic energies are converted to thermal ones.
Indeed, the deceleration and deflection of ions result in an increase of local ion temperatures. We find the ion temperature at the central high density region can reach as high as $400$ eV. 
The sharp boundary between high density (temperature) and low density (temperature) plasmas indicate a strong shock is formed in this colliding process.
When a high density and high temperature plasma is formed, one would expect an expansion at latter time. 
At $t=333$ ps, the directions of ion velocities point from the central spot to the outwards, which do indicate an expansion.
In reality, the highest ion temperature is not located at the central high density region as one would expect. 
As shown in Fig.\ \ref{fig6}(e), the ion temperature at the shock fronts can be as high as $600$ eV. 
This feature of ion temperature distribution indicates that kinetic effects play roles in this strong shock \cite{kinetic_shock1,kinetic_shock2,kinetic_shock3,kinetic_shock4}.
In Fig.\ \ref{fig6}(g)-(i), the evolution of magnetic fields is also displayed.  Self-generated electromagnetic fields as high as $1000$ T are produced in such collisions. 
 
In order to depict the heating and overlapping of colliding plasma jets more clearly, in Fig.\ \ref{fig7}, we have displayed the ion particles in phase-space plots.
Fig.\ \ref{fig7}(a)-(c) show the $z-v_z$, $v_z-v_y$ and $v_x-v_y$ plots of ions at $t=100$ ps. 
As one can see, at $z=100$ $\mu$m, i.e., the place where two plasma jets meet, 
those ions at initial colliding velocity of $300$ km/s start merging with each other; the width of velocity distribution is increased and a local quasi-Maxwellian velocity distribution is therefore formed. 
Fig.\ \ref{fig7}(d)-(f) show the $z-v_z$, $v_z-v_y$ and $v_x-v_y$ plots of ions at $t=166$ ps.
As one can see, at this stage, almost the entire ion particles have merged with each other and a global quasi-Maxwellian velocity distribution is formed. 

In Fig.\ \ref{fig8}, the temporal evolutions of global ion kinetic energy and electron kinetic energy are displayed, with the total simulation time over $300$ ps. 
Solid lines display the results for two degenerate plasma jets with initial density of 100 g/cc, temperature of 65 eV and drifting velocity of 300 km/s.
At $t=50$ ps, the two plasma jets start to collide with each other, converting kinetic energy from ions to electrons.
At $t=160$ ps, the kinetic energy of ions reaches the lowest value and on the contrary, the kinetic energy of electrons reaches the highest value. 
This should be the stagnation stage where the density compression is the highest and the plasma temperature at the central region also the highest. 
The lasting time of the stagnation stage is about $50$ ps, as marked by the black dashed lines in the plot. In the DCI scheme, 
following the head-on collision is a rapid heating by energetic electrons. As the rapid heating occurs within $10$ ps, 
it is apparently that a $50$ ps time-window for the rapid heating is well enough.
After the stagnation stage, the free expansion of the plasma follows, with a conversion of kinetic energies from electrons to ions.
From the temporal evolution, one can quantitatively calculate the energy conversion efficiency from the colliding kinetic energy to thermal ones.
It is reasonable to assume the electron temperature and the ion temperature are the same. 
The increase of electron kinetic energy as marked with  $\delta$ and bounded by red dashed lines is equivalent to the decrease of ion kinetic energy as bounded by blue dashed lines.
The conversion efficiency from the colliding kinetic energy to thermal ones can therefore be calculated as $\eta=2\delta/k_i|_{t=0}$, which is $85\%$ in this studied case. 
Dashed lines display the results of a toy model in which electrons are treated as classical particles as a comparison. In this toy model, electrons are also initialized as Fermi-Dirac distributions to ensure the same initial condition, however Pauli-exclusion is excluded in the following evolutions. We therefore see a rapid energy conversion from electrons to ions at the very beginning for the toy model, which would finally lead to a lower conversion efficiency.

In a recent experimental measurement \cite{dci2}, with the total driving laser energy only of 10 kJ, the density of the supersonic plasma jet can reach as high as $5.5\sim8$ g/cm$^3$ and the average velocity of the jet is about $135$ km/s. 
During colliding, a
stagnation phase lasts about $\sim200$ ps is formed, and the maximum density of the plasma core is increased to $46\pm24$ g/cm$^3$, with a density increasement of $5\sim8$.
By analyzing the velocity and temperature before and after colliding, it is found that $89.5\%$ of the kinetic energy
is converted into thermal energy.
Although the plasma conditions are somewhat different, both the predicted values of confinement time at stagnation and energy conversion ratio agree with experimental measurement at a reasonable range. 
Considering in the experiment two colliding plasam jets are converging, this 3D effect would bring additional density increasement, which might explain why the density increasement is much higher in experimental measurement than predicted by 2D simulations with a flat configuration. In the following, realistic configurations with 3D effects are planned along with a large range of colliding parameters.

\section{Conclusions}
The head-on collision of high density plasma jets is of key importance for the DCI scheme, and also finds great applications in many other high energy density researches. 
To overcome the simulation limitations that appear in the fluid description, in this paper, we propose a new simulation method for large-scale high density plasmas with an ingenious kinetic-ion and kinetic/hydrodynamic-electron treatment. This method takes advantages of modern particle-in-cell simulation techniques and binary Monte Carlo collisions, which include both long-range collective electromagnetic fields and short-range particle-particle interactions. 
Especially, in this method, the restrictions of simulation grid size and time step, that appear in a fully kinetic description, can be eliminated. 
Furthermore, the collisional coupling and state-dependent coefficients, that are usually approximately used in a fluid description, are no longer required. The correctness and robustness of the new simulation method are verified, by comparing with fully kinetic simulations at small scales and purely hydrodynamic
simulations at large scale. Following the conceptual design of the DCI scheme, the colliding process of two plasma jets with initial density of 100 g/cc, initial thermal temperature of 70 eV, and counter-propagating velocity at 300 km/s is investigated using this new simulation method. Quantitative values, including density increment, increased plasma temperature, confinement time at stagnation and conversion efficiency from the colliding kinetic energy to thermal energy, are obtained with a density increment of about three times, plasma temperature of 400 eV, confinement time at stagnation of $50$ ps and conversion efficiency of $85\%$. These values agree with the recent experimental measurements at a reasonable range and might serve as a reference for future detailed studies. 
This simulation approach might also find great applications in (laboratory) astrophysics, and plasma blow-off inside hohlraums in indirect-drive ICF researches.  

\begin{acknowledgments}
We acknowledge Professor Xiantu He, Zhengming Sheng, Yanyun Ma, Weimin Wang and Han Xu for useful discussion on the physical scenarios.
This work was supported by the Strategic Priority Research Program of Chinese Academy of Sciences (Grant No. XDA250050500), National Natural Science Foundation of China (Grant No. 12075204) and Shanghai Municipal Science and Technology Key Project (No. 22JC1401500). D. Wu thanks the sponsorship from Yangyang Development Fund.
\end{acknowledgments}

\section{Data Availability Statement}
The data that supports the findings of this study are available within the article.


\begin{thebibliography}{99} 

\bibitem{icf1} J. Nuckolls, L. Wood, A. Thiessen, G. Zimmerman, ``Laser compression of matter to super-high densities'', Nature 239, 139 (1972).
\bibitem{icf2} J. Lindl, RL. McCrory, EM. Campbell, ``Progress toward ignition and propagating burn in inertial confinement fusion'', Phys. Today 45, 32 (1992).
\bibitem{icf3} X. T. He, J. W. Li, Z. F. Fan, L. F. Wang, J. Liu, K. Lan, J. F. Wu, and W. H. Ye, ``A hybrid-drive nonisobaric-ignition scheme for inertial confinement fusion''
Phys. Plasmas 23, 082706 (2016).
\bibitem{icf4} J. Zhang, W. M. Wang, X. H. Yang, D. Wu, Y. Y. Ma, J. L. Jiao, Z. Zhang, F. Y. Wu, X. H. Yuan, Y. T. Li, and J. Q. Zhu, ``Double-cone ignition scheme for inertial confinement fusion'', 
A Phil. Trans. R. Soc. A 13, 378 (2020).
\bibitem{hol1} L. Q. Shan,H. B. Cai, W. S. Zhang, Q. Tang, F. Zhang, Z. F. Song, B. Bi, F. J. Ge, J. B. Chen, D. X. Liu, W. W. Wang, Z. H. Yang, W. Qi, C. Tian, Z. Q. Yuan, 
B. Zhang, L. Yang, J. L. Jiao, B. Cui, W. M. Zhou, L. F. Cao, C. T. Zhou, Y. Q. Gu, B. H. Zhang, S. P. Zhu, and X. T. He,
``Experimental evidence of kinetic effects in indirect-drive inertial confinement fusion hohlraums'', Phys. Rev. Lett. 120, 195001 (2018).
\bibitem{hol2} L. B. Hopkins, N. Meezan, S. Le Pape, L. Divol, A. Mackinnon, D. Ho, M. Hohenberger, O. Jones, G. Kyrala, J. Milovich et al., ``First high-convergence cryogenic implosion in a near-vacuum hohlraum'', Phys. Rev. Lett. 114, 175001 (2015).
\bibitem{pic} C. K. Birdsall, A. B. Langdon, ``Plasma physics via computer simulation'', Taylor and Francis, New York, 2005.
\bibitem{fl1} D. P. Higginson, P. Amendt, N. Meezan, W. Riedel, H. G. Rinderknecht, S. C. Wilks, and G. Zimmerman, ``Hybrid particle-in-cell simulations of laser-driven plasma interpenetration, heating, and entrainment'', Phys. Plasmas 26, 112107 (2019).
\bibitem{fl2} H. Xu, X. H. Yang, J. Liu, and M. Borghesi, ``Control of fast electron propagation in foam target by high-Z doping'', Plasma Phys. Control. Fusion 61, 025010 (2019).
\bibitem{fl3} C. Thoma, D. R. Welch, R. E. Clark, D. V. Rose, and I. E. Golovkin, ``Hybrid-PIC modeling of laser-plasma interactions and hot electron generation in gold hohlraum walls'', Phys. Plasmas 24, 062707 (2017).
\bibitem{fl4} Hong-bo Cai, Xin-xin Yan, Pei-lin Yao, and Shao-ping Zhu, ``Hybrid fluid-particle modeling of shock-driven hydrodynamic instabilities in a plasma'', Matter Radiat. Extremes 6, 035901 (2021).
\bibitem{pauli1} A. E. Turrell, M. Sherlock, and S. J. Rose, ``A Monte Carlo algorithm for degenerate plasmas'', J. Comput. Phys. 249, 13 (2013).
\bibitem{pauli2} D. Wu, W. Yu, S. Fritzsche, and X. T. He, ``Particle-in-cell simulation method for macroscopic degenerate plasmas'', Phys. Rev. E 102, 033312 (2020). 
\bibitem{Taki} T. Takizuka, H. Abe, J. Comput. ``A binary collision model for plasma simulation with a particle code'', Phys. 25 205 (1977).
\bibitem{Nanbu} K. Nanbu, S. Yonemura, ``Weighted Particles in Coulomb Collision Simulations Based on the Theory of a Cumulative Scattering Angle'', Journal of Computational Physics  145 639 (1998).
\bibitem{Sentoku} Y. Sentoku, A. J. Kemp, ``Numerical methods for particle simulations at extreme
densities and temperatures: Weighted particles, relativistic collisions and reduced currents'', Journal of Computational Physics 227, 6846 (2008).
\bibitem{lapins1} D. Wu, W. Yu, S. Fritzsche, and X. T. He, ``A high-order implicit particle-in-cell method for plasma simulations at solid densities'', Phys. Rev. E 100, 013207 (2019).
\bibitem{lapins2} D. Wu, W. Yu, Y. T. Zhao, D. H. H. Hoffmann, S. Fritzsche, and X. T. He, ``Particle-in-cell simulation of transport and energy deposition of intense proton beams in solid-state materials'', 
Phys. Rev. E 100, 013208 (2019).
\bibitem{multi} Fuyuan Wu, Rafael Ramis, Zhenghong Li, Yanyun Chu, Jianlun Yang, Zhen Wang, Shijiang Meng, Zhanchang Huang and Jiaming Ning, ``Numerical Simulation of the Interaction Between Z-Pinch Plasma and Foam Converter Using Code MULTI ($\#18353$)'', Fusion Science and Technology, 72, 726 (2017).
\bibitem{kinetic_shock1} Bellei C, Amendt P A. Shock-induced mix across an ideal interface, Phys. Plasmas, 24, 040703 (2017).
\bibitem{kinetic_shock2} Amendt P, Bellei C, Ross J S, et al. Ion separation effects in mixed-species ablators for inertial-confinement-fusion implosions, Phys. Rev. E, , 91, 023103 (2015).
\bibitem{kinetic_shock3} Rinderknecht H G, Rosenberg M J, Li C K, et al. Ion thermal decoupling and species separation in shock-driven implosions, Phys. Rev. Letters, 114, 025001 (2015). 
\bibitem{kinetic_shock4} Hua R, Kim J, Sherlock M, et al. Self-generated magnetic and electric fields at a Mach-6 shock front in a low density helium gas by dual-angle proton radiography, Physical Review Lett., 123, 215001 (2019).

\bibitem{dci2} Zhe Zhang, Xiao-Hui Yuan, Yi-Hang Zhang, Hao Liu, Ke Fang, Cheng-Long Zhang, Zheng-Dong Liu, Xu Zhao, Quan-Li Dong, Gao-Yang Liu, Yu Dai, Hao-Chen Gu, Yu-Tong Li, Jian Zheng, Jia-Yong Zhong, Jie Zhang, ``Efficient energy transition from kinetic to internal energy in supersonic collision of high-density plasma jets from conical implosions'', Acta  Phys.  Sin.  71, 155201 (2022).
\end{thebibliography}
\end{document}